\documentclass[iop,apj]{emulateapj}

\newcommand{\kms}{{\mathrm{km~s}^{-1}}}

\shorttitle{Classical Cepheids in the Nuclear Stellar Disk}
\shortauthors{Matsunaga et al.}

\begin{document}

\title{Kinematics of classical Cepheids in the Nuclear Stellar Disk}

\author{
Noriyuki Matsunaga\altaffilmark{1},
Kei Fukue\altaffilmark{1},
Ryo Yamamoto\altaffilmark{2},
Naoto Kobayashi\altaffilmark{2,3},
Laura Inno\altaffilmark{4,5},
Katia Genovali\altaffilmark{5},
Giuseppe Bono\altaffilmark{5},
Junichi Baba\altaffilmark{6},
Michiko~S. Fujii\altaffilmark{7},
Sohei Kondo\altaffilmark{8},
Yuji Ikeda\altaffilmark{8},
Satoshi Hamano\altaffilmark{2},
Shogo Nishiyama\altaffilmark{9},
Tetsuya Nagata\altaffilmark{10},
Wako Aoki\altaffilmark{7},
and
Takuji Tsujimoto\altaffilmark{7}
}
\altaffiltext{1}{Department of Astronomy, School of Science, The University of Tokyo, 7-3-1 Hongo, Bunkyo-ku, Tokyo 113-0033, Japan; matsunaga@astron.s.u-tokyo.ac.jp}
\altaffiltext{2}{Institute of Astronomy, School of Science, The University of Tokyo, 2-21-1 Osawa, Mitaka, Tokyo 181-0015, Japan}
\altaffiltext{3}{Kiso Observatory, Institute of Astronomy, School of Science, The University of Tokyo, 10762-30 Mitake, Kiso-machi, Kiso-gun, Nagano 397-0101, Japan}
\altaffiltext{4}{European Southern Observatory, Karl-Schwarzschild-Str.~2, 85748 Garching bei M\"{u}nchen, Germany} 
\altaffiltext{5}{Dipartimento di Fisica, Universit\'{a} di Roma Tor Vergata, Via della Ricerca Scientifica 1, 00133 Rome, Italy}
\altaffiltext{6}{Earth-Life Science Institute, Tokyo Institute of Technology, 2-12-1 Ookayama, Meguro-ku, Tokyo 152-8551, Japan}
\altaffiltext{7}{National Astronomical Observatory of Japan, 2-21-1 Osawa, Mitaka, Tokyo 181-8588, Japan}
\altaffiltext{8}{Koyama Astronomical Observatory, Kyoto Sangyo University, Motoyama, Kamigamo, Kita-ku, Kyoto 603-8555, Japan}
\altaffiltext{9}{Miyagi University of Education, 149 Aramaki-aza-Aoba, Aoba-ku, Sendai, Miyagi 980-0845, Japan}
\altaffiltext{10}{Department of Astronomy, Kyoto University, Kitashirakawa-Oiwake-cho, Sakyo-ku, Kyoto 606-8502, Japan}

\begin{abstract}
Classical Cepheids are useful tracers of the Galactic young
stellar population because their
distances and ages can be determined from their period-luminosity
and period-age relations.
In addition, the radial velocities and chemical abundance of the Cepheids
can be derived from spectroscopic observations, 
providing further insights into
the structure and evolution of the Galaxy.
Here, we report the radial velocities of classical Cepheids
near the Galactic Center, three of which
were reported in 2011, the other reported for the first time.
The velocities of these Cepheids suggest that the stars orbit
within the Nuclear Stellar Disk, a group of stars and interstellar matter
occupying a region of $\sim 200$~pc around the Center,
although the three-dimensional velocities cannot be determined until
the proper motions are known.
According to our simulation, these four Cepheids formed
within the Nuclear Stellar Disk like younger stars and
stellar clusters therein.
\end{abstract}


\section{Introduction}

The Galactic Center comprises
complex systems of stars and interstellar matter (gas and dust)
displaying interesting and important phenomena
\citep[e.g., see the review by][]{Morris-1996}.
The Galactic bulge, extending over a few kilo-parsecs, is
populated predominantly by old stars, roughly 10~Gyr old
\citep{Zoccali-2003}, and possibly by a smattering of younger stars
\citep[a few Gyr old;][]{Bensby-2013}.
In contrast, $\sim 200$-pc region around Sgr~A$^*$ at the Center
hosts many young stars and is the region of current star formation
\citep{Serabyn-1996}.
These young stars are distributed throughout a disk-like structure 
called the Nuclear Stellar Disk
\citep[hereafter NSD;][]{Launhardt-2002}.
Interstellar gas and dust
occupy the so-called the Central Molecular Zone
\citep[CMZ;][]{Morris-1996}
found in the same region.

The most prominent stars in the NSD are young luminous stars
such as O/B-type stars and Wolf-Rayet stars.
These are found in large numbers in  three stellar clusters;
Arches, Quintuplet, and
the nuclear star cluster surrounding Sgr~A*.
The Arches and Quintuplet clusters are at most
6~Myr old \citep{Stolte-2008,Liermann-2012,Schneider-2014},
and young massive stars also reside in the nuclear star cluster
\citep{Krabbe-1995}.
Similarly, young and massive stars are found across the NSD field
\citep{YusefZadeh-2009,Mauerhan-2010}.
Recently \citet{Matsunaga-2011} discovered 
classical Cepheids in the direction of the NSD, and their 
distances (estimated from period-luminosity relations) are
consistent with the distance to the NSD.
From the period-age relation \citep{Bono-2005},
these Cepheids were found to be similarly aged ($\sim 25$~Myr).
Populations of relatively young stars are also evidenced by
the concentration of OH/IR stars and mass-losing red supergiants
\citep{Lindqvist-1992}.
\citet{Blommaert-1998} found that some OH/IR stars
are very luminous, suggesting ages of
$\sim 1$~Gyr or even $\sim 100$~Myr
\citep[see also][]{vanLoon-2003}.
However, these estimates have large uncertainties because we lack
understanding of the mass-loss phenomena in the stars
\citep{Gallart-2005}.
From surface-density distributions, \citet{Matsunaga-2013} suggested that 
a fraction of old stellar population represented by short-period Miras 
belongs to the NSD, although NSD membership requires confirmation by 
kinematical information.

The disk-like structure of the NSD and CMZ suggests that
these systems spin around the Galactic Center.
In fact, the $l$-$v$ diagram reveals elliptical orbits of the molecular gas
in this region \citep{Binney-1991,Stark-2004},
which are understood as $x_2$ orbits expected near or inside
the inner Lindblad resonance \citep{Contopoulos-1980,Athanassoula-1992}.
On the other hand, $x_1$ orbits are elongated
along the bar potential and are exhibited by a portion of gas
within the Galaxy.
Gas transfer from $x_1$ to $x_2$ orbits is considered to supply
interstellar matter to the CMZ
and to sustain star formation therein \citep{Stark-2004}.
Numerical simulations conducted by \citet{Kim-2011} demonstrated that
gas moving along $x_1$
orbits can fall down to $x_2$ orbits, forming
a ring in which stars form.

Some observational data show that stars in the NSD also orbit
in a manner somewhat consistent with the $x_2$ orbits.
The first evidence of such rotation came from observations of
maser emission by OH/IR stars and
large-amplitude variables \citep{Lindqvist-1992,Deguchi-2004}.
The radial velocities of the massive clusters Arches and Quintuplet
(approximately $+100~\kms$ at positive Galactic longitudes)
also suggest prograde rotation in the NSD
\citep{Figer-2002,Liermann-2009}.
However, \citet{Stolte-2008,Stolte-2014} 
reported that the proper motions of Arches and Quintuplet clusters are
too large for movement along closed $x_2$ orbits,
but may be consistent with transitional trajectories
from $x_1$ to $x_2$ orbits.
To investigate the evolution of the NSD, 
we require the kinematics of stars of various ages;
however, previously studied objects in the NSD are limited.

Here we report spectroscopic observations of classical Cepheids 
within $0.4^\circ$ of the Center.
These stars are young ($\sim 25$~Myr), already suggesting that they belong
to the NSD rather than to the extended bulge dominated by old stars,
and their rotations should appear in the NSD. 
The dynamical (rotational) time scale in the NSD is several Myr;
hence, these stars may have orbited several times since their birth. 
Such tracers would provide important insight into
the formation and dynamical evolution of stars in the NSD.
The pulsations of Cepheids cause red- and blue-shifts
of their stellar absorption lines,
and spectra should preferably be taken at different pulsation epochs
\citep[see, for instance,][]{Marconi-2013}.
In addition, the severe interstellar extinction toward the NSD demands infrared
observations, although little research exists on the
infrared spectroscopy of Cepheids
\citep{Sasselov-1990,Sabbey-1995,Nardetto-2011}.
The present work utilizes the $H$-band spectra between
{15680~\AA} and {17870~\AA}; in this wavelength range
the extinction for the Cepheids found near the NSD is around 4.5~mag
in contrast to more than 30~mag in the optical.
These infrared spectra are used for measuring the radial velocities
in order to confirm the membership to the NSD and to study the kinematics
within the NSD.

\section{Observations}

\subsection{Targets}

\begin{figure}
\begin{center}
\includegraphics[clip,width=0.98\hsize]{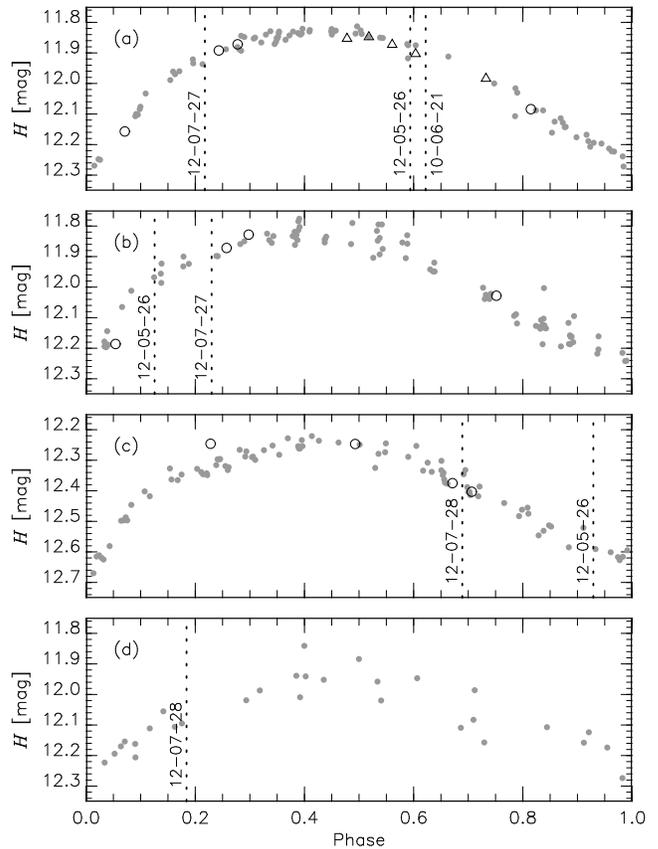}
\end{center}
\caption{
$H$-band light curves of the target Cepheids.
Magnitudes of {GCC-a}, -b, and -c
in June 2010 and July 2012 are indicated
by open triangles and open circles, respectively.
The photometric data in \citet{Matsunaga-2013} are also included.
Vertical dotted lines indicate the
dates (YY-MM-DD) and phases of the IRCS spectroscopic observations
(see Tables~\ref{tab:log} and \ref{tab:vel}).
\label{fig:LCs}}
\end{figure}

Our targets are the three Cepheids reported in
\citet{Matsunaga-2011,Matsunaga-2013}, named {GCC-a}, -b, and -c,
and a similar newly-discovered Cepheid, named {GCC-d} (Table~\ref{tab:obj}).
Our recent survey of variable stars detected this
Cepheid at $0.32^\circ$ from Sgr~A$^*$. In this survey, 
142 fields-of-view were captured
along the Galactic plane ($|l|<10^\circ$, $b=0^\circ$)
by the SIRIUS camera
attached to the Infrared Survey Facility (IRSF).
The monitoring started in April 2007 and continued until May 2012;
the numbers of phases was typically 30.
The exposure time was 8~s per visit, $\sim 6$ times shorter than
that in \citet{Matsunaga-2011}.
Among several classical Cepheids discovered by our group
(Matsunaga {et~al.}, in prep),
the coordinates of {GCC-d} and its distance from the Sun
places this star near the NSD. The period of {GCC-d},
determined in the same way as in \citet{Matsunaga-2013},
approximates those of the other three Cepheids, 
suggesting a similar age ($\sim 25$~Myr) based on 
the period-age relation \citep{Bono-2005}.
The photometric measurements 
and the light curve of {GCC-d} are presented in
Table~\ref{tab:GCCd} and
Figure~\ref{fig:LCs}, respectively.
Barycentric Julian Date (BJD) in the Barycentric Dynamical Time standard
was calculated for each $JHK_{\rm s}$ dataset \citep{Eastman-2010},
while Modified Julian Date (MJD)
is also listed to make the comparison with previous publications easier.

Also at the IRSF, we conducted additional photometry
of {GCC-a}, -b, and -c in June 2010 and July 2012
(Table~\ref{tab:newJHK}; and see also Figure~\ref{fig:LCs}).
By combining the new data with the previous data,
we obtained more precise periods (Table~\ref{tab:obj}) with a longer baseline,
and can more accurately estimate the pulsational phases 
at which the following spectroscopic observations were made.

\begin{deluxetable}{cccrrr}
\tablewidth{0pt}
\tablecaption{Target list\label{tab:obj}}
\tablehead{
\colhead{Object} & \colhead{RA} & \colhead{Dec}
& \colhead{$l$} & \colhead{$b$} & \colhead{Period} \\
\colhead{} & \colhead{(J2000.0)} &\colhead{(J2000.0)} &\colhead{(deg)} &\colhead{(deg)} &\colhead{(days)} 
}
\startdata
{GCC-a} & 17:46:06.01 & $-$28:46:55.1 & $+$0.186 & $-$0.009 & 23.528 \\
{GCC-b} & 17:45:32.27 & $-$29:02:55.2 & $-$0.105 & $-$0.043 & 19.942 \\
{GCC-c} & 17:45:30.89 & $-$29:03:10.5 & $-$0.112 & $-$0.041 & 22.755 \\
{GCC-d} & 17:44:56.90 & $-$29:13:33.7 & $-$0.324 & $-$0.026 & 18.886 
\enddata
\end{deluxetable}

\begin{deluxetable*}{cccccccc}
\tablewidth{0pt}
\tablecaption{Photometric catalogue of {GCC-d}\label{tab:GCCd}}
\tablehead{
\colhead{BJD} & \colhead{MJD} &
\colhead{$J$} & \colhead{$e_{J}$} &  
\colhead{$H$} & \colhead{$e_{H}$} &
\colhead{$K_{\rm s}$} & \colhead{$e_{K_{\rm s}}$} \\ 
\colhead{} & \colhead{} & \colhead{(mag)} & \colhead{(mag)} & \colhead{(mag)} & \colhead{(mag)} & \colhead{(mag)} & \colhead{(mag)} 
}
\startdata
2454172.62180 & 54172.12153 & 15.63 & 0.05 & 12.10 & 0.01 & 10.33 & 0.03 \\
2454220.48339 & 54219.97875 & 15.71 & 0.11 & 12.08 & 0.01 & 10.27 & 0.01 \\
2454224.47362 & 54223.96870 & 15.87 & 0.08 & 12.12 & 0.01 & 10.33 & 0.01 \\
2454231.50812 & 54231.00275 & 15.35 & 0.07 & 12.02 & 0.01 & 10.25 & 0.02 \\
2454246.56109 & 54246.05498 & 15.97 & 0.07 & 12.21 & 0.01 & 10.41 & 0.04 \\
2454284.32459 & 54283.81819 & 15.71 & 0.06 & 12.16 & 0.01 & 10.38 & 0.02 \\
2454321.38376 & 54320.87922 & 15.75 & 0.08 & 12.19 & 0.01 & 10.41 & 0.01 \\
2454326.40654 & 54325.90239 & 15.49 & 0.09 & 11.99 & 0.01 & 10.21 & 0.01 \\
2454349.36209 & 54348.85997 & 15.55 & 0.05 & 11.96 & 0.01 & 10.16 & 0.03 \\
2454360.36648 & 54359.86543 &  ...  & ...  & 12.11 & 0.04 & 10.45 & 0.03 \\
2454384.32315 & 54383.82442 &  ...  & ...  & 11.94 & 0.01 & 10.19 & 0.01 \\
2454385.26846 & 54384.76981 & 15.44 & 0.11 & 11.95 & 0.02 & 10.17 & 0.01 \\
2454537.57722 & 54537.07699 & 15.51 & 0.06 & 11.88 & 0.01 & 10.12 & 0.01 \\
2454573.51165 & 54573.00799 & 15.54 & 0.07 & 11.94 & 0.02 & 10.18 & 0.03 \\
2454619.63155 & 54619.12519 & 15.70 & 0.08 & 12.11 & 0.01 & 10.28 & 0.01 \\
2454697.25949 & 54696.75580 & 15.80 & 0.09 & 12.17 & 0.01 & 10.39 & 0.02 \\
2454715.33246 & 54714.83041 &  ...  & ...  & 12.16 & 0.01 & 10.35 & 0.01 \\
2454743.28785 & 54742.78852 &  ...  & ...  & 12.01 & 0.02 & 10.16 & 0.01 \\
2454889.64600 & 54889.14703 &  ...  & ...  & 12.05 & 0.01 & 10.29 & 0.01 \\
2454919.63460 & 54919.13269 &  ...  & ...  & 12.16 & 0.01 & 10.25 & 0.01 \\
2455010.50274 & 55009.99624 & 15.87 & 0.09 & 12.02 & 0.01 & 10.22 & 0.01 \\
2455020.38676 & 55019.88049 & 15.50 & 0.07 & 12.17 & 0.02 & 10.34 & 0.01 \\
2455058.28738 & 55057.78332 & 15.85 & 0.07 & 12.15 & 0.02 & 10.36 & 0.02 \\
2455114.24664 & 55113.74783 &  ...  & ...  & 12.22 & 0.01 & 10.40 & 0.01 \\
2455324.42128 & 55323.91606 & 15.64 & 0.07 & 12.11 & 0.01 & 10.29 & 0.02 \\
2455434.33571 & 55433.83256 & 15.66 & 0.07 & 12.27 & 0.01 & 10.45 & 0.01 \\
2455617.60239 & 55617.10364 &  ...  & ...  & 12.11 & 0.02 & 10.46 & 0.06 \\
2455725.51261 & 55725.00603 & 15.26 & 0.05 & 11.84 & 0.02 & 10.09 & 0.05 \\
2456050.47590 & 56049.97095 &  ...  & ...  & 11.95 & 0.02 & 10.21 & 0.01 \\
2456052.46567 & 56051.96059 &  ...  & ...  & 11.99 & 0.02 & 10.23 & 0.02 
\enddata
\end{deluxetable*}

\begin{deluxetable*}{ccccccccc}
\tablewidth{0pt}
\tablecaption{Additional photometry in $JHK_{\rm s}$ for {GCC-a}, -b, and -c\label{tab:newJHK}}
\tablehead{
\colhead{Object} &
\colhead{BJD} & \colhead{MJD} &
\colhead{$J$} & \colhead{$e_{J}$} &  
\colhead{$H$} & \colhead{$e_{H}$} &
\colhead{$K_{\rm s}$} & \colhead{$e_{K_{\rm s}}$} \\ 
\colhead{} & \colhead{} & \colhead{} & \colhead{(mag)} & \colhead{(mag)} & \colhead{(mag)} & \colhead{(mag)} & \colhead{(mag)} & \colhead{(mag)} 
}
\startdata
GCC-a&2455365.52974&55365.02314 & 15.45 & 0.03 & 11.85 & 0.01 &  9.95 & 0.01 \\
     &2455366.47019&55365.96359 & 15.47 & 0.03 & 11.85 & 0.01 &  9.96 & 0.01 \\
     &2455367.47477&55366.96817 & 15.47 & 0.03 & 11.87 & 0.01 &  9.99 & 0.01 \\
     &2455368.48445&55367.97786 & 15.56 & 0.02 & 11.90 & 0.01 & 10.01 & 0.01 \\
     &2455371.51335&55371.00677 &  ...  & ...  & 11.98 & 0.01 & 10.11 & 0.01 \\
     &2456126.34383&56125.83790 & 15.79 & 0.04 & 12.08 & 0.01 & 10.22 & 0.01 \\
     &2456132.37312&56131.86750 & 15.69 & 0.06 & 12.16 & 0.01 & 10.30 & 0.01 \\
     &2456136.42675&56135.92137 & 15.48 & 0.06 & 11.89 & 0.01 & 10.04 & 0.01 \\
     &2456137.23325&56136.72792 & 15.48 & 0.04 & 11.87 & 0.01 & 10.02 & 0.01 \\
\hline
GCC-b&2456126.34712&56125.84120 & 15.53 & 0.04 & 12.03 & 0.04 & 10.21 & 0.03 \\
     &2456132.37648&56131.87087 & 15.54 & 0.05 & 12.19 & 0.03 & 10.31 & 0.04 \\
     &2456136.43006&56135.92469 & 15.34 & 0.05 & 11.87 & 0.02 & 10.04 & 0.03 \\
     &2456137.23654&56136.73122 & 15.25 & 0.04 & 11.83 & 0.02 &  9.97 & 0.03 \\
\hline
GCC-c&2456126.34712&56125.84120 & 16.12 & 0.06 & 12.25 & 0.02 & 10.19 & 0.01 \\
     &2456132.37648&56131.87087 & 16.33 & 0.11 & 12.25 & 0.01 & 10.13 & 0.01 \\
     &2456136.43006&56135.92469 &  ...  & ...  & 12.38 & 0.01 & 10.24 & 0.01 \\
     &2456137.23654&56136.73122 &  ...  & ...  & 12.40 & 0.01 & 10.28 & 0.01 
\enddata
\end{deluxetable*}

\subsection{Spectroscopy with Subaru/IRCS}

In 2010 and 2012, we collected near-infrared spectra of our Cepheids at
several epochs when possible using  
the Infrared Camera and Spectrograph (IRCS) attached
to the Subaru 8.2~m telescope \citep{Kobayashi-2000}.
This instrument allows us to obtain high-resolution
($\lambda/\Delta\lambda = 20,000$) echelle
spectra in the near-infrared region. We observed {GCC-a} three times,
{GCC-b} and -c twice, and {GCC-d} once. Most spectra were in the $H$-band
but one $K$-band spectrum of {GCC-a} was taken in June 2010
(Table~\ref{tab:log}). During the July 2012 run, we used
the adoptive optics system, AO188, with laser star guiding \citep{Hayano-2010}.
Unfortunately, the AO-guided observations 
in the other runs (June 2010 and May 2012)
were interrupted by instrumental problems and poor seeing condition,
and the quality of the spectra was poorer than that in July 2012.
Nevertheless, all data were usable in the radial velocity measurements.

\begin{deluxetable*}{cccccr}
\tablewidth{0pt}
\tablecaption{Log of spectroscopic observations\label{tab:log}}
\tablehead{
\colhead{UTC\tablenotemark{a}} &
\colhead{BJD\tablenotemark{b}} &
\colhead{Object} & 
\colhead{Band} &
\colhead{Integrations\tablenotemark{c}} &
\colhead{S/N\tablenotemark{d}} \\
\colhead{} & \colhead{} & \colhead{} & \colhead{} &
\colhead{$t_1~(^{\rm s}) \times N$} & \colhead{} 
}
\startdata
2010-06-21~09:30&2455368.90243&{GCC-a}&$K$&$300 \times  8$ & 45 \\
2012-05-26~14:00&2456074.08954&{GCC-a}&$H$&$300 \times 12$ & 25 \\
2012-05-26~11:00&2456073.96453&{GCC-b}&$H$&$300 \times 12$ & 25 \\
2012-05-26~12:40&2456074.03398&{GCC-c}&$H$&$300 \times 12$ & 25 \\
2012-07-27~07:30&2456135.81792&{GCC-a}&$H$&$300 \times 12$ & 100 \\
2012-07-27~09:00&2456135.88041&{GCC-b}&$H$&$300 \times 12$ & 100 \\
2012-07-28~08:00&2456136.83868&{GCC-c}&$H$&$300 \times 12$ & 100 \\
2012-07-28~10:40&2456136.94978&{GCC-d}&$H$&$300 \times  8$ & 85 
\enddata
\tablenotetext{a}{Coordinated Universal Time at around the middle of the observation}
\tablenotetext{b}{Barycentric Julian Date calculated from the UTC}
\tablenotetext{c}{Duration of each integration ($t_1$) and the number $N$}
\tablenotetext{d}{Typical signal-to-noise ratio at the continuum level}
\end{deluxetable*}

\section{Analysis}

\begin{figure}
\begin{center}
\includegraphics[clip,width=0.98\hsize]{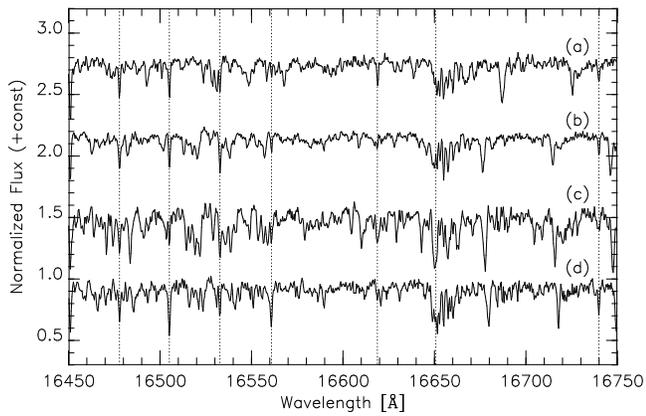}
\end{center}
\caption{
Detailed plot of the Subaru/IRCS spectra in the $H$-band
for four Cepheids (data collected in July 2012). 
Vertical dotted lines indicate telluric lines used for
wavelength calibration. Intrinsic absorption lines 
show different redshifts of the targets.
\label{fig:comp2}}
\end{figure}

Data reduction of the spectra was performed by standard procedures,
including background subtraction, wavelength calibration, and normalization
using the Image Reduction and Analysis Facility
(IRAF\footnote{IRAF is distributed by the National Optical Astronomy
Observatory, which is operated by the Association of Universities for
Research in Astronomy(AURA)
under cooperative agreement with the National Science Foundation.}).
Wavelength calibration was based on the telluric absorption lines within
individual target spectra. Spectra of each object in each run
were combined into a single spectrum and normalized by 
the continuum level which can be reasonably traced.
We retained the telluric absorption lines for redshift measurements,
as explained below. Detailed plots
of the target Cepheid spectra are compared in 
Figure~\ref{fig:comp2}.
Clear differences are observed in the redshifts of the four objects.

\begin{figure}
\begin{center}
\includegraphics[clip,width=0.98\hsize]{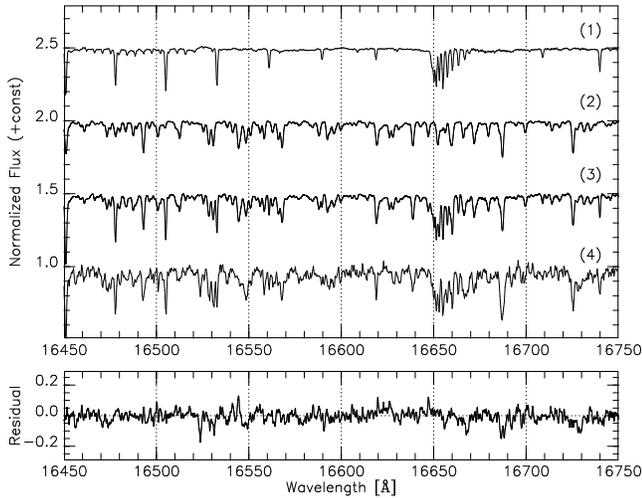}
\end{center}
\caption{
Illustration of the method used to
measure the radial velocities. 
A telluric spectrum constructed from an A-type star~(1)
was convolved with a model spectrum shifted by a trial velocity~(2)
to construct a synthesized spectrum~(3). This spectrum is
compared with the observed spectrum~(4).
The lower panel shows the difference between the spectra (3) and (4).
\label{fig:comp}}
\end{figure}

To estimate the radial velocities, we compared
the observed spectra with synthetic spectra 
constructed by the tools ATLAS9 and SYNTHE developed by
\citet{Kurucz-1993}\footnote{http://kurucz.harvard.edu/}.
They calculate plane-parallel atmospheric models and generate
the synthetic spectra for a given stellar parameter set
including the effective temperature $T_{\rm eff}$ and the gravity $\log g$.
These tools do not take the pulsation into account,
but we assume the validity of the quasi-static approximation
for the Cepheid atmosphere \citep[see, e.g.,][]{Molinaro-2011}.
Line lists were extracted from the Vienna Atomic Line Database (VALD).
Although these lists differ from the original list 
provided by Kurucz, the redshift estimates were essentially unaffected
(the following results in our paper would remain unchanged within $1~\kms$).
Figure~\ref{fig:comp} illustrates the method
used to measure the redshifts;
a telluric absorption spectrum constructed from an A-type star observation~(1)
was convolved with a model spectrum shifted by a trial redshift~(2)
to construct a synthesized spectrum~(3). The synthetic spectrum was then
compared with the observed spectrum~(4).
Each target spectrum was assigned the redshift that minimized
the $\chi ^2$ value of the difference between (3) and (4);
the residual obtained by subtracting (3) from (4) is indicated
in the lower panel of Figure~\ref{fig:comp}.
The reduced $\chi^2$ value is typically $\sim 25$.
Although some differences exist among the spectra of the four Cepheids 
and between the model and observed spectra,
such differences were annulled by the many common absorption lines,
thus exerting negligible effects on the radial velocity measurements.
Therefore, for every spectrum, we applied
the stellar model spectrum with the following parameter set: 
$T_\mathrm{eff}=5000$~K, $\log g=1.0$, and $Z=Z_\odot$.
We applied this method to each of the following ranges in five echelle orders:
17500--17870~\AA ~in the order $H32$, 16960--17340~\AA ~in $H33$,
16460--16840~\AA ~in $H34$, 15980--16150~\AA ~in $H35$, and 15680--15840~\AA
~in $H36$.
In case of the $K$-band spectrum of GCC-a,
the following five ranges are used:
23920--24480~\AA ~in $K23$, 
22920--23480~\AA ~in $K24$, 
22000--22560~\AA ~in $K25$, 
21160--21680~\AA ~in $K26$, 
and 20460--20780~\AA ~in $K27$.
The telluric absorption lines distributed in these ranges ensure
well-calibrated wavelength scales.
These spectral ranges yielded
consistent velocities (within a few $\kms$;
see $\sigma _V$ in Table~\ref{tab:vel}),
and the statistical error was estimated from the scattering in the measured
velocities over the five ranges.
The obtained velocities were then transformed into
barycentric velocities, $V_{\rm bary}$,
and velocities relative to
the Local Standard of Rest (LSR), $V_{\rm LSR}$, assuming
the standard solar motion \citep{Crovisier-1978,Reid-2009}.

Cepheid pulsations can alter the radial velocities measured at
each epoch by tens of $\kms$ \citep[e.g.,][]{Pejcha-2012}.
First of all, we need the pulsational phase in each spectrum
to discuss such variations.
The pulsational phase is usually defined based on
the optical $V$-band curves, but it is not possible 
to obtain the optical light curves of our targets
because of severe extinction.
Here, we define the zero phase at the $H$-band minimum.
For our purpose, we require only the relative phase offsets between
the $H$-band light curves and the radial velocity curves.

\begin{figure}
\begin{center}
\includegraphics[clip,width=0.98\hsize]{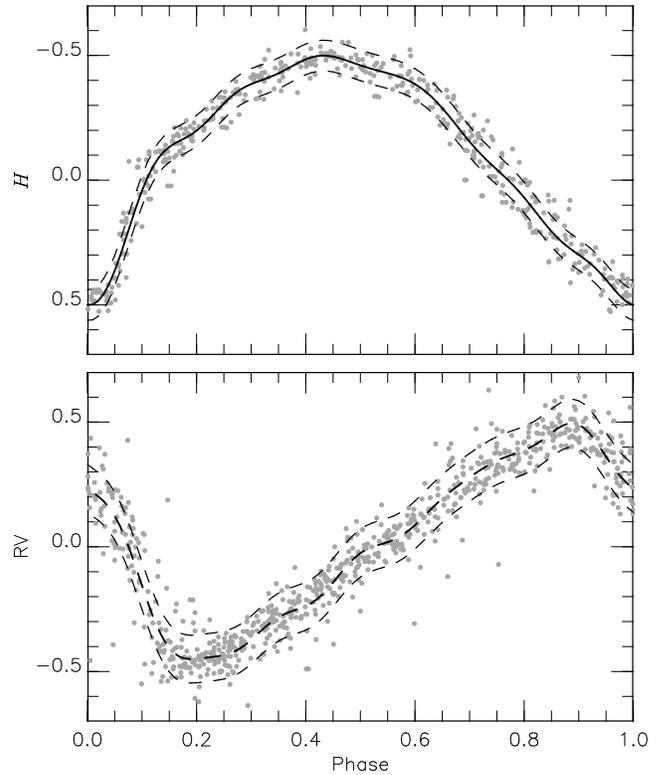}
\end{center}
\caption{
Normalized variations of $H$-band light curves and radial velocity
curves for nearby Cepheids with $P\simeq 20$~days \citep{Groenewegen-2013}.
Fitted template curves and the $\pm 1\sigma$ ranges are indicated
by solid and dashed curves, respectively.
\label{fig:fit_combined}}
\end{figure}
\begin{figure}
\begin{center}
\includegraphics[clip,width=0.98\hsize]{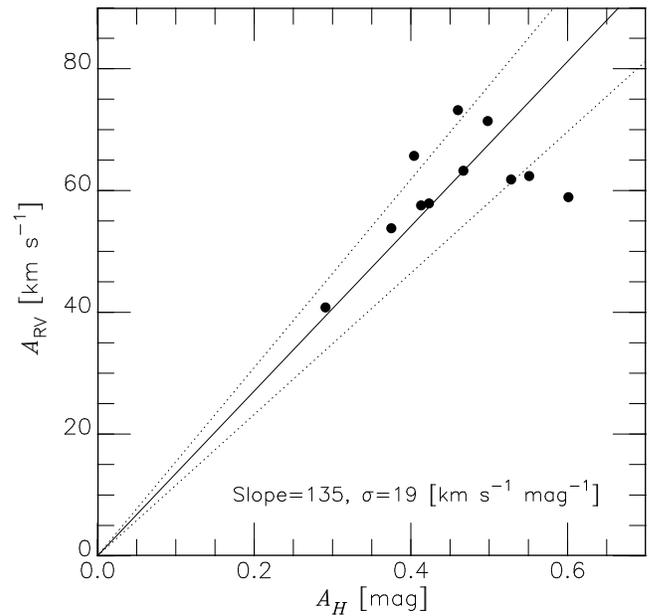}
\end{center}
\caption{
Velocity amplitude versus amplitude of $H$-band light curves
for nearby Cepheids with $P\simeq 20$~days \citep{Groenewegen-2013}.
Object names and dataset used are stated in the text.
The slope of the solid line was obtained as the average
of $A_{\rm RV}/A_{H}$ values, and its standard deviation $\sigma$ is
indicated by the dotted lines. 
\label{fig:comp_amp}}
\end{figure}

To obtain mean
velocities of the target Cepheid, we first constructed a template of
the velocity variations of the Cepheids with $P\simeq 20$~days.
All our targets have similar periods, but the prediction template
is yet to be established.
We adopted the infrared photometry and radial velocities
of 11 nearby Cepheids with $18 \leq P \leq 24$~days
(Table~\ref{tab:templates})
compiled by \citet{Groenewegen-2013}.
First, the $H$-band and velocity curves of each template Cepheid were
fitted by seventh-order Fourier series,
and the amplitude and the mean were obtained.  
Here we refer to the peak-to-peak amplitude of a Fourier series fit and
the average of its maximum and minimum
as the amplitude and the mean (or zero), respectively.
Then, amplitude-normalized templates of the $H$-band and velocity curves 
were constructed combining the data of the 11 Cepheids
(Figure~\ref{fig:fit_combined}).
Within the period range, the template
Cepheids show similar variations in both $H$-band light and velocity. 
In addition,
the amplitudes of the template $H$-band light curves are well correlated
with those of the velocity curves, as shown in Figure~\ref{fig:comp_amp}.
The amplitude ratio, $A_{\rm RV}/A_{H}=135~\kms~{\rm mag}^{-1}$, is taken
as the mean value for the 11 template stars,
and the standard deviation is $19~\kms~{\rm mag}^{-1}$.
We can use this ratio to predict velocity amplitudes based on
$H$-band light curves.

\begin{deluxetable}{crrrrrrr}
\tablewidth{0pt}
\tablecaption{The Cepheids considered for the templates\label{tab:templates}}
\tablehead{
\colhead{Object} & \colhead{Period} & \colhead{$A_{H}$} & \colhead{$A_{\rm RV}$} \\
\colhead{} & \colhead{(days)} & \colhead{(mag)} & \colhead{($\kms$)} 
}
\startdata
YZ~Aur & 18.193 & 0.38 & 54 \\
VY~Car & 18.905 & 0.40 & 66 \\
RU~Sct & 19.703 & 0.42 & 58 \\
VX~Cyg & 20.133 & 0.41 & 58 \\
RY~Sco & 20.320 & 0.29 & 41 \\
RZ~Vel & 20.397 & 0.46 & 73 \\
WZ~Sgr & 21.851 & 0.47 & 63 \\
BM~Per & 22.952 & 0.60 & 59 \\
WZ~Car & 23.013 & 0.55 & 62 \\
VZ~Pup & 23.175 & 0.50 & 71 \\
SW~Vel & 23.428 & 0.53 & 62 
\enddata
\tablecomments{
The peak-to-peak amplitudes of $H$-band light curves ($A_{H}$)
and those of velocity curves ($A_{\rm RV}$) are obtained
by fitting seventh-order Fourier series. References to the original
photometric and spectroscopic data are listed in Table~1 of
\citet{Groenewegen-2013}.
}
\end{deluxetable}

\begin{figure}
\begin{center}
\includegraphics[clip,width=0.98\hsize]{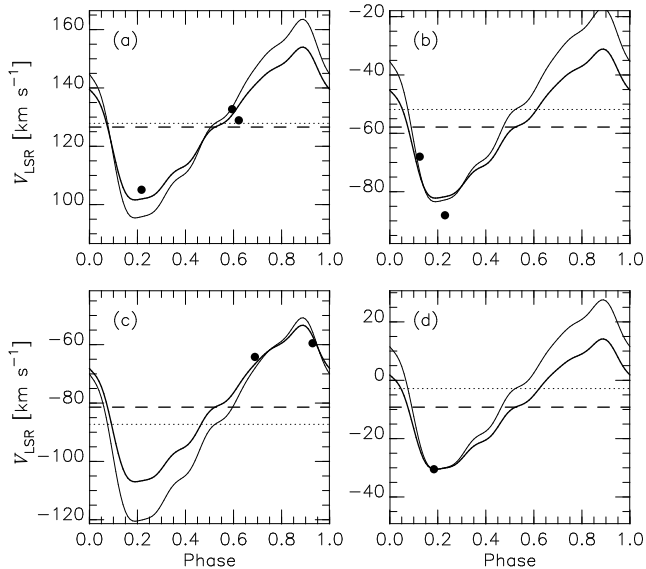}
\end{center}
\caption{
Velocities at individual epochs indicated by filled circles are compared
with the velocity curve templates. Panels (a), (b), (c), and (d) show 
the data for our targets {GCC-a}, -b, -c, and -d, respectively.
Thick curves have the amplitudes obtained based on the amplitude ratio 
$A_{\rm RV}/A_{H}=135~\kms~{\rm mag}^{-1}$, while
thin curves have those with the ratio larger by 30~\%
\citep[][see text]{Nardetto-2011}.
Horizontal lines indicate the estimated mean velocities;
thick dashed lines are given by the smaller amplitude ratio
and dotted lines by the larger ratio.
\label{fig:phase_cor}}
\end{figure}

We estimated the $H$-band amplitudes and the pulsation
phases of our target Cepheids in individual spectroscopic observations
by fitting seventh-order Fourier series to the light curves.
The velocity amplitude $A_{\rm RV}$
was estimated combining the $A_{H}$ and the amplitude ratio,
and the mean velocity was obtained by shifting the velocity curve template
multiplied by the predicted amplitude. When two or more measurements
were available, the estimated mean velocities of each object were averaged.
Figure~\ref{fig:phase_cor} shows the template, which is vertically shifted
to fit the measured values.
Note that the amplitudes are predicted based on the amplitude ratio
$A_{\rm RV}/A_{H}$, rather than by fitting
the black circles in Figure~\ref{fig:phase_cor}.
Nevertheless, the predicted velocity curves adequately accommodate
the measured velocities, particularly for {GCC-a} and -c.
The estimated mean
velocities are listed in Table~\ref{tab:vel} and
indicated by horizontal lines in Figure~\ref{fig:phase_cor}.

\begin{deluxetable*}{crrrrrrr}
\tablewidth{0pt}
\tablecaption{Measured radial velocities of the Cepheids\label{tab:vel}}
\tablehead{
\colhead{Object} & \colhead{BJD} & \colhead{Phase} &  
\colhead{$V_\mathrm{bary}$} & \colhead{$V_\mathrm{LSR}$} &
\colhead{$\Delta V$} & \colhead{$\sigma_{V}$} \\
\colhead{} & \colhead{} & \colhead{} & \colhead{($\kms$)} & \colhead{($\kms$)} & \colhead{($\kms$)} & \colhead{($\kms$)}
}
\startdata
{GCC-a}& 2455368.90243 & 0.62 & $+121.1$ & $+128.9$ & $-7$ & 2.1 \\
       & 2456074.08954 & 0.59 & $+125.0$ & $+132.7$ & $-4$ & 1.9 \\
       & 2456135.81792 & 0.22 & $ +97.3$ & $+105.1$ & $+25$ & 1.4 \\
       &               & Mean & $ +119$ & $+127$ &      \\
\hline
{GCC-b}& 2456073.96453 & 0.13 & $ -75.6$ & $ -68.0$ & $+17$ & 6.2 \\
       & 2456135.88041 & 0.23 & $ -95.8$ & $ -88.1$ & $+24$ & 0.8 \\
       &               & Mean & $ -66$ & $-58$ &      \\
\hline
{GCC-c}& 2456074.03398 & 0.93 & $ -67.2$ & $ -59.5$ & $-24$ & 2.5 \\
       & 2456136.83868 & 0.69 & $ -71.8$ & $ -64.2$ & $-15$ & 1.4 \\
       &               & Mean & $ -89$ & $-81$ &      \\
\hline
{GCC-d}& 2456136.94978 & 0.18 & $ -38.1$ & $ -30.5$ & $+21$ & 0.8 \\
       &               & Mean & $ -17$ & $-9$ &      
\enddata
\tablecomments{
The barycentric velocities ($V_\mathrm{bary}$) and
velocities relative to the LSR ($V_\mathrm{LSR}$)
are listed at individual epochs; each velocity was calculated from
measurements over five echelle orders 
and the standard error $\sigma_{V}$ of the measurements is also listed.
Velocities for each object are averaged
after correcting for the pulsational effects ($\Delta V$).
}
\end{deluxetable*}

The uncertainty in predicting the amplitude ratio, $A_{\rm RV}/A_{H}$,
introduces the largest error in our correction of the pulsation effect.
First, the standard deviation in Figure~\ref{fig:comp_amp} leads to
an error of $7.6~\kms$ in predicted velocity amplitude
considering that the $H$-band amplitudes of our Cepheids are
approximately 0.4~mag. In addition, there can be a systematic difference
between the velocity amplitudes from optical spectroscopic data
and the counterparts from infrared data.
While the $A_{\rm RV}$ for the temperate Cepheids were obtained
with optical spectra, our spectra for the target Cepheids are in the infrared.
\citet{Nardetto-2011} found that the velocity amplitude of $\ell$~Car
($P=35.56$~d)
from infrared, $K$-band, absorption lines is 1.3 times larger than 
that from optical absorption lines based on their 4-epoch spectroscopic
observations using CRIRES at VLT. If this is true for 
our target Cepheids with $A_{H}\sim 0.4$~mag, we should consider
the velocity amplitude $70~\kms$ rather than $55~\kms$.
The difference, if any, between the velocity curves
from optical, $H$- and $K$-band absorption lines is not established.
Unfortunately, our spectroscopic observations are too few to
determine the amplitudes.
Thin curves in Figure~\ref{fig:phase_cor}
indicate fits obtained by assuming the larger amplitude ratio.
Two horizontal lines in each panel indicate how the estimated mean
velocity is affected by the amplitude ratio.
The effect is small in the case of GCC-a which have two measurements
around the mean, and the three points slightly prefer the predicted curve
with the smaller amplitude ratio.
The mean velocities of the others get offset by $\sim 7~\kms$.
Velocity curves of Cepheids obtained with infrared spectra
and their difference from those obtained with optical spectra should
be investigated with more data as \citet{Nardetto-2011} also suggested.
We here adopt the results with the smaller amplitude ratio
and they are listed in Table~\ref{tab:vel}.
Considering the above uncertainties, the velocity amplitudes
are expected be between $\sim 45~\kms$ 
and $\sim 80~\kms$ for a Cepheid with $A_{H}=0.4$~mag.
Since we adopted $55~\kms$, this range introduces an error of $\pm 13~\kms$
in mean velocity,
because the mean estimate is affected by half the change of
the velocity amplitude if a measured velocity is obtained at
the extreme.

\section{Discussion}

\begin{figure*}
\begin{center}
\includegraphics[clip,width=0.60\hsize]{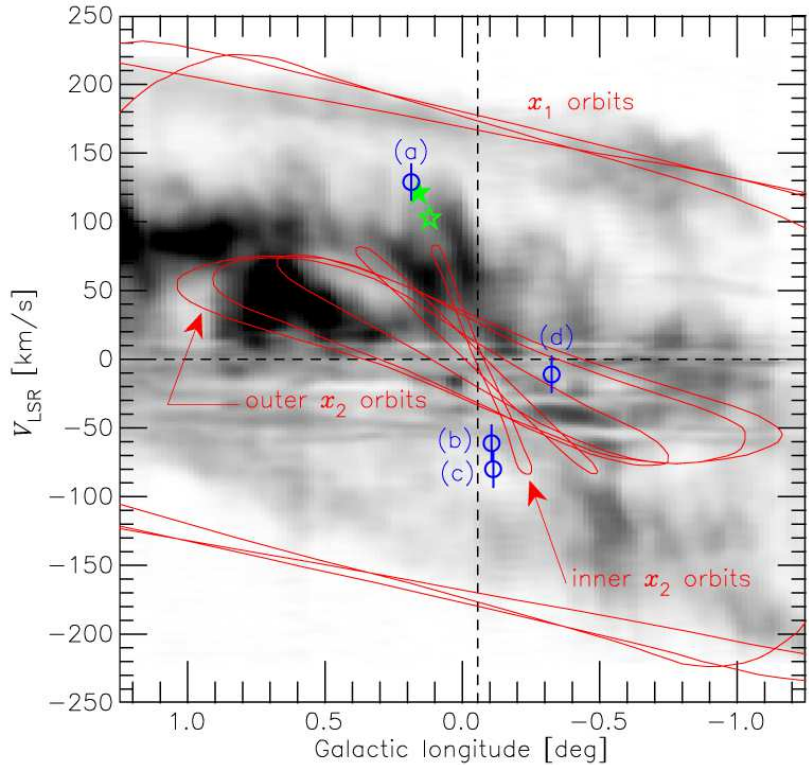}
\end{center}
\caption{
Velocities relative to the LSR ($V_\mathrm{LSR}$)
of our four Cepheids are plotted against
the Galactic longitude (open circles).
The error bars have the size of $13~\kms$ which is discussed in the text.
The velocities and locations of
the Arches and Quintuplet clusters 
are indicated by open and filled star symbols, respectively.
The $x_1$ and $x_2$ orbits calculated by
\citet{Bissantz-2003} are taken from Figure~1 of \citet{Stark-2004}
and indicated by solid curves.
The background displays the $l$-$v$ diagram of the CO~$J=1-0$ emissions
taken from \citet{Oka-1998}. 
(A color version of this figure is available in the online journal.)
\label{fig:LV}}
\end{figure*}

Figure~\ref{fig:LV} plots the $V_\mathrm{LSR}$ of the Cepheids
against the Galactic longitude.
The longitude and $V_\mathrm{LSR}$ of {GCC-a} are both positive,
whereas those of {GCC-b}, -c, and -d are negative.
This result is consistent with the prediction that the Cepheids
orbit around Sgr~A$^*$ similar to other objects in the NSD.
The solid curves in Figure~\ref{fig:LV} are extracted from
Figure~1 in \citet{Stark-2004} and indicate several $x_1$ and $x_2$ orbits
from the model of \citet{Bissantz-2003}.
Their model adopts the bar gravitational potential
constructed from photometric data collected by the {\it COBE} satellite
\citep{Bissantz-2002}.

Examining the $l$-$v$ diagram, we find that
{GCC-b} and -c follow inner $x_2$ orbits, whereas
{GCC-d} follows an outer $x_2$ orbit.
{GCC-b} and -c are close to each other not only in projected position but
also in radial velocity. The error of their estimated distances
is $\pm 500$~pc, which does not allow us to determine their line-of-sight
positions within the NSD (radius $\sim 200$~pc).
It is tempting to argue that the two stars
could have been formed in the same star-forming event, but
their proper motions and 3D velocities are required
to confirm the similarity in the kinematics and to further investigate
their possible association.

In contrast, the velocity $V_{\rm LSR}$ of {GCC-a} (127~$\kms$),
is larger than
the expected velocities for $x_2$ orbits ($< 80 \kms$) at its
projected distance of 30~pc from Sgr~A$^*$.
The $x_2$ orbits depend on
the mass distribution within the NSD, which is dominated
by stars rather than the central supermassive black hole, except
at the very center \citep{Launhardt-2002}.
The mass enclosed within a circular orbit of a radius 30~pc
and velocity of $130~\kms$ is $2.5\times 10^8~{\rm M_\odot}$,
more than twice that predicted by the mass distribution 
constructed from infrared photometric data
\citep{Bissantz-2002,Launhardt-2002}.
The kinematics of OH/IR stars and SiO masers are consistent
with the smaller mass \citep{Lindqvist-1992,Deguchi-2004}.
On the other hand, the location and velocity of {GCC-a} approximates
those of the Arches and Quintuplet clusters,
although the two clusters, $\sim 4$~Myr, are much younger
than GCC-a, $\sim 25$~Myr.
The radial velocities of these clusters, plotted in Figure~\ref{fig:LV},
are taken from \citet{Figer-2002} and \citet{Liermann-2009}.
It is known that these clusters have large proper motions and
do not follow closed $x_2$ orbits
\citep{Stolte-2008,Stolte-2014}.
Stolte {et~al.} suggested that these clusters are following
transitional trajectories between $x_1$ and $x_2$ orbits.
Although the proper motion is necessary to compare their 3D kinematics,
{GCC-a} may have followed a motion similar to
those of the two clusters when it was young
and may not yet have settled into a circular orbit.
Notably, \citet{Lindqvist-1992}
and \citet{Deguchi-2004} reported objects with similarly large velocities, and 
large-velocity molecular clouds are evidenced in the CO emission 
in the background of Figure~\ref{fig:LV} \citep{Oka-1998}.
In summary, despite the unavailability of proper motions,
we can infer 
that the four Cepheids are rotating within the NSD, 
although not necessarily along closed orbits.

According to our results, the Cepheids are not only currently located
within the NSD but most likely formed therein.
To investigate the birth sites of NSD stars,
we investigated their dynamical histories in
a N-body/SPH simulation of the Galaxy (Baba et al.~in prep.). 
This simulation self-consistently accounts for the stellar dynamics,
self-gravity of the gas, 
radiative cooling, heating by interstellar far-ultraviolet radiation, 
energy feedback from supernova explosions, and HII regions.
In several simulation runs, we identified star particles aged 20--30 Myr
in the NSD
and traced their motions to 100 Myr ago, when they existed as gas clouds.
At $\sim 75$~Myr before star formation, i.e. during the gas cloud phase, 
some of the tracked particles had reached the CMZ while others were
moving along the bar. 
However, all these clouds were trapped within the CMZ at the time 
of star formation. Star formation was scarce along the bar 
but common within the CMZ, consistent with the simulations 
of \citet{Kim-2011}.

\section{Summary}

We conducted infrared spectroscopic observations of classical Cepheids
near the Galactic Center.
From the measured radial velocities of the Cepheids,
we inferred that they orbit within the NSD in a manner similar to
other contained objects. Preliminary
simulation suggested that these Cepheids formed in the same disk region
20--30~Myr ago and may have undergone several rotations,
rather than becoming trapped in the disk after birth.
The orbits of the Cepheids should be further investigated
once the proper motions
become available, and a detailed study on their kinematics would
provide important clues on stellar formation and dynamical evolution
in the NSD region.
Furthermore, by investigating the chemical abundance of these objects,
we could elucidate the 
chemical evolution in and around this region;
this topic is left for future study.

\acknowledgments

We acknowledge the anonymous referee for many useful comments
to improve the manuscript.
We express our thanks to Takahiro Naoi, Nagisa Oi, and Tatsuhito Yoshikawa
who collected additional IRSF photometric data for our targets.
We are also grateful to Subaru support astronomers,
Miki Ishii, Yosuke Minowa, and Tae-Soo Pyo,
for their help during our Subaru observations in 2010 and 2012.
Martin Groenewegen provided the dataset
for nearby template Cepheids which were published in 2013.
A part of this paper was written in a hospital in Cape Town, South Africa,
where the first author NM was treated after his accidental fall 
in the IRSF dome on May 3, 2014. NM is grateful to everyone, 
in both South Africa and Japan, who rescued and supported him
after the accident;
especially to Patricia Whitelock, John Menzies, Michael
and Connie Feast, 
and Rie Matsunaga for their warm support.
This work has been supported by Grants-in-Aid for Scientific Research
(No.~22840008, 23684005, and 26287028)
from the Japan Society for the Promotion of Science (JSPS).

{\it Facilities:} \facility{IRSF}, \facility{Subaru}

\end{document}